\documentclass[]{spie}  %>>> use for US letter paper
\usepackage[]{graphicx}
\usepackage[]{amsmath}
\usepackage{float}
\usepackage{array}
\usepackage{multirow}
\newcommand{\PreserveBackslash}[1]{\let\temp=\\#1\let\\=\temp}
\newcolumntype{C}[1]{>{\PreserveBackslash\centering}p{#1}}
\newcolumntype{R}[1]{>{\PreserveBackslash\raggedleft}p{#1}}
\newcolumntype{L}[1]{>{\PreserveBackslash\raggedright}p{#1}}
\title{Pseudo Dual Energy CT Imaging using Deep Learning Based Framework: Initial Study}

\author{Sui~Li}
\author{Yongbo~Wang}
\author{Yuting~Liao}
\author{Ji~He}
\author{Dong~Zeng}
\author{Zhaoying~Bian}
\author{Jianhua~Ma}
\affil{Department of Biomedical Engineering, Southern Medical
University, Guangzhou, Guangdong 510515, China}
\affil{Guangzhou
Key Laboratory of Medical Radiation Imaging and Detection
Technology, Guangzhou 510515, China}
\authorinfo{Correspondence: D.Z.: E-mail: zd1989@smu.edu.cn}
\begin{document}
  \maketitle

%%%%%%%%%%%%%%%%%%%%%%%%%%%%%%%%%%%%%%%%%%%%%%%%%%%%%%%%%%%%%
\begin{abstract}
Dual energy computed tomography (DECT) has become of particular
interest in clinic recent years.  The DECT scan comprises two
images, corresponding to two photon attenuation coefficients maps
of the objects.  Meanwhile, the DECT images are less accessible
sometimes, compared to the conventional single energy CT (SECT).
This motivates us to simulate pseudo DECT (pDECT) images from the
SECT images.  Inspired by recent advances in deep learning, we
present a deep learning based framework to yield pDECT images from
SECT images, utilizing the intrinsic characteristics underlying
DECT images, i.e., global correlation and high similarity. To
demonstrate the performance of the deep learning based framework,
a cascade deep ConvNet (CD-ConvNet) approach is specifically
presented in the deep learning framework. In the training step,
the CD-ConvNet is designed to learn the non-linear mapping from
the measured energy-specific (i.e., low-energy) CT images to the
desired energy-specific (i.e., high-energy) CT images.  In the
testing step, the trained CD-ConvNet can be used to yield desired
high-energy CT images from the low-energy CT images, and then
produce accurate basic material maps.  Clinical patient data were
employed to validate and evaluate the presented CD-ConvNet
approach performance.  Both visual and quantitative results
demonstrate the presented CD-ConvNet approach can yield high
quality pDECT images and basic material maps.
\end{abstract}

%>>>> Include a list of keywords after the abstract


\section{METHODOLOGY}

\subsection{CD-ConvNet architecture in pDECT imaging}
%The multienergy CT images can be considered as a spectral sequence
%of spatial imges, i.e.,
%\begin{equation}
%\label{eq3} X = \left\{ {{x_i},{\text{ }}i \leq   M} \right\},
%\end{equation}

%\noindent where $X$ denotes the MECT images, $i$ corresponds to
%one of $M$ energy bins.

\begin{figure}[H]
\centerline{\includegraphics[scale=0.8]{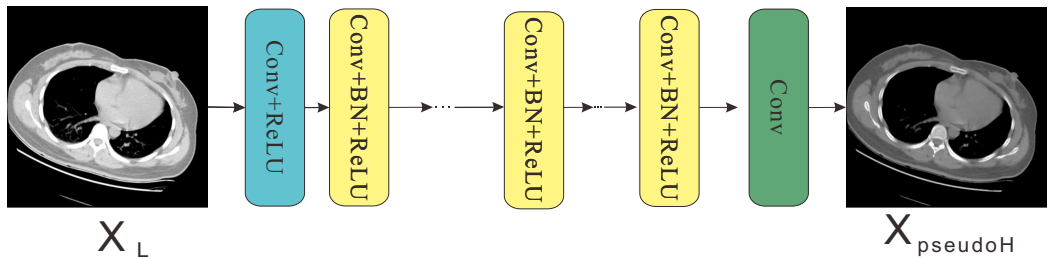}} \caption{The
architecture of used ConvNet.}\label{fig1}
\end{figure}

The structure of the ConvNet\cite{Lecun2015} is shown in
Fig.~\ref{fig1}.  It consisted of sequential convolution modules
which are convolution (Conv), batch normalization (BN) and ReLU
layers.

% To include a figure: 1

From the Fig.~\ref{fig1}, it can be seen that the first layer of
ConvNet is excluded BN layers and the last module contained only
convolution layer.  In the training, $3\times3\times64$
convolution kernels were used in all the convolution layers except
that the first layer which contains $7\times7\times64$ convolution
kernels in convolution layer, and there are 20 modules in a single
ConvNet totally.

\begin{figure}[H]
\centerline{\includegraphics[scale=0.8]{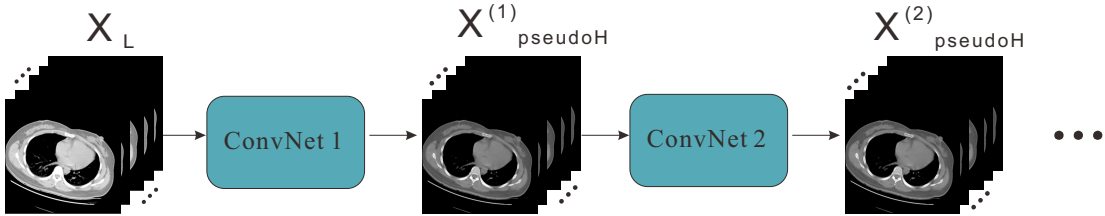}} \caption{The
structure of our cascade deep convolutional neural
network.}\label{fig2}
\end{figure}

Although one ConvNet is intuitionistic for SECT images compound to
pseudo DECT images, the image quality might not be satisfactory. A
ConvNet with large enough receptive field is needed to fully
extract the useful feature information from the low-energy CT
images. However, such a ConvNet might need more training samples
and longer time to learn excessive number of parameters.
Therefore, to address this issue, we presented a cascade deep
ConvNet (CD-ConvNet) for the deep learning framework to improve
the image quality of pseudo DECT images.  The structure of the
CD-ConvNet is shown in Fig.~\ref{fig2}.

% To include a figure: 2

In Fig.~\ref{fig2}, ${X_{L}}$ and ${X_{PredictH}}$ denote the
low-energy CT images and the pseudo high-energy CT images,
respectively. The first ConvNet was trained to map ${X_{L}}$ to
high-energy CT images, then the ${X_{L}}$ was predicted to get the
${X_{PseudoH}^{(1)}}$. Then a cascaded ConvNet was trained to map
${X_{PseudoH}^{(1)}}$ to high-energy CT images in the similar way
to get the ${X_{PseudoH}^{(2)}}$. The following cascaded was
constructed in the same way with the previous one, as showed in
Fig.~\ref{fig2}.
%%-----------------------------------------------------------
\subsection{Implementation details}

The clinical datasets were used for this experiment, and it
contains abdominal spectral CT images collected from 8 patients
with two energy spectra.  In order to increase sample data, we
simulated both of the two energy CT images with different
rotations of each slice.  We choose 7 patients datasets images as
the training datasets of each cascade and the remaining one
patient data was used as testing dataset.  There were 2,121 slices
in total in the training datasets and 334 slices in the testing
datasets. The presented cascade deep convolution neural network
(CD-ConvNet) was implemented using the MatConvNet toolbox in
MATLAB (2015b) environment.  Less than 10 hours were cost for
training one single ConvNet on a workstation equipped with a GPU
(Nvidia Quadro M4000) with 4 GB RAM.

The training of the ConvNet in the CD-ConvNet is to determine a
set of parameters i.e, $\theta =\left\{{{W_d},{B_d};d =
1,2,\cdots,D}\right\}$ , via minimizing the loss between the
network pseudo high-energy image ${X_{PseudoH}}$ and the reference
high-energy CT image ${X_{H}}$. Given a set of
pairs$\left\{{X_{PseudoH}},{X_{H}}\right\}$, a commonly used loss
function for regression tasks, is defined as:

\begin{equation}
\left( \theta  \right) = \frac{1}{N}\mathop \sum \limits_{i =1}^N
{\left\| {{X_{PseudoH}} - {X_{H}}} \right\|^2},
\end{equation}

\noindent where ${N}$ is the number of training samples.

The loss function was minimized via mini-batch stochastic gradient
descent (SGD) algorithm.  The batch size, momentum, and weight
decay for the min-batch SGD were set to 128, 0.7, and $10^{-4}$,
respectively.  The learning rate of all the convolution layers set
to $10^{-3}$ during the first 5 epochs and $10^{-5}$  in the rest
epochs. The filter weights of each layer were initialized with a
Gaussian function with a zero mean and standard deviation of
${\sqrt{{2}/{M}}}$, with ${M}$ indicating the number of incoming
nodes of one neuron. The initial biases of each convolution layer
were set to 0.

\section{RESULTS}

\begin{figure}[H]
\centerline{\includegraphics[scale=0.8]{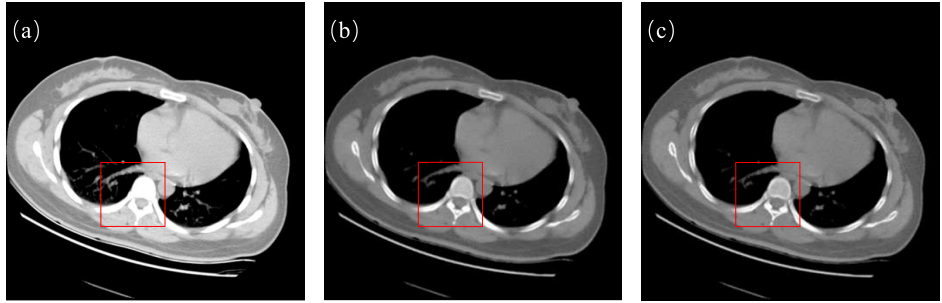}}
\caption{Selected single energy image from (a) low-energy CT image
(b) pseudo high-energy CT image from CD-ConvNet and (c) reference
high-energy CT image.  All the images are displayed with the same
window: [0.011 0.024] $mm^{-1}$.}\label{fig3}
\end{figure}

\begin{figure}[H]
\centerline{\includegraphics[scale=0.8]{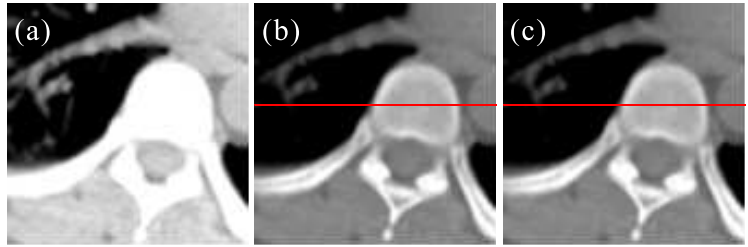}}
\caption{Zoomed images indicated by the red rectangle in
Fig.~\ref{fig3}. (a) low-energy CT image (b) pseudo high-energy CT
image from CD-ConvNet (c) reference high-energy CT image.  All the
images are displayed with the same window: [0.011 0.024]
}\label{fig4}
\end{figure}

\begin{figure}[H]
\centerline{\includegraphics[scale=0.6]{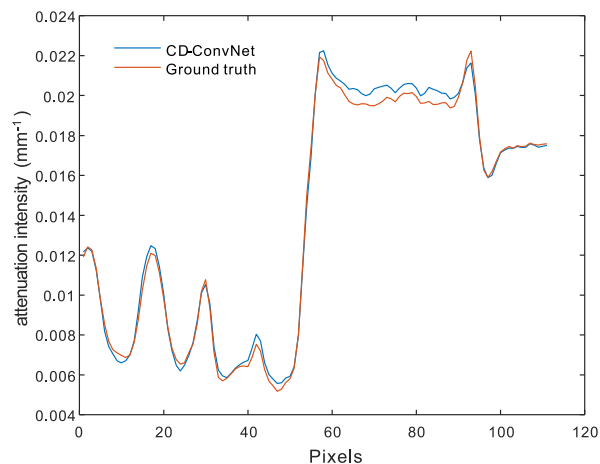}}
\caption{Profiles of pseudo high-energy CT image ${(CD-ConvNet)}$
and reference high-energy CT image. }\label{fig5}
\end{figure}

Fig.~\ref{fig3} presents the the low-energy CT image, high-energy
CT image, and pseudo high-energy CT images reconstructed by the
CD-ConvNet. It can be observed that the pseudo high-energy CT
image from CD-ConvNet is close to the reference high-energy CT
image in visual inspection.
%The difference between the pseudo
%high-energy CT image and the input low-energy CT image is shown in
%Fig.~\ref{fig5}(a) and the difference between reference
%high-energy CT image and pseudo high-energy CT image of the
%selected slice is shown in Fig.~\ref{fig5}(b). It can be seen that
%the values in difference map between the two high-energy CT images
%are almost close to zero.
Moreover, Fig.~\ref{fig4} shows the zoom-in-view of the region of
interest (ROI) indicated by red rectangle in Fig.~\ref{fig3}.
Fig.~\ref{fig5} depicts the line profiles along the red line as
indicated in Fig.~\ref{fig4}, it can be seen that the profile from
the pseudo high-energy CT image from presented CD-ConvNet agrees
closely with profile from reference high-energy CT image. The
qualitative experiments can demonstrate that the CD-ConvNet can
produce accurate pseudo high-energy CT image from low-energy CT
image visually.

% To include a figure: 3

% To include a figure: 4

% To include a figure: 5

\begin{figure}[H]
\centerline{\includegraphics[scale=0.5]{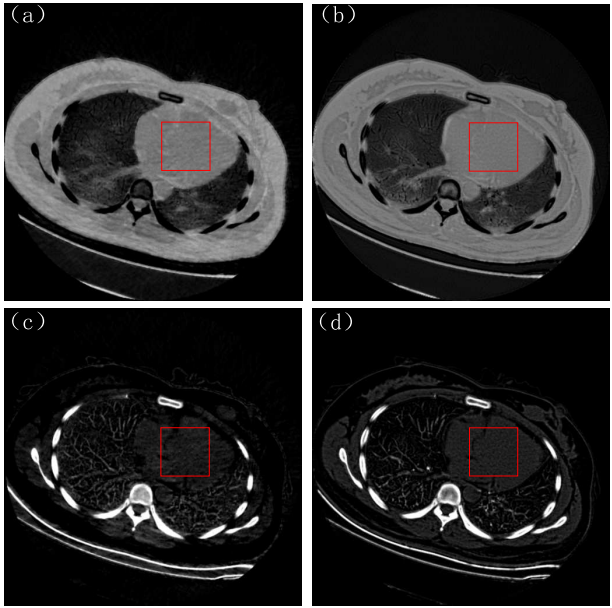}} \caption{
Material decompositions from DECT CT image and pDECT image. The
first row is foe water equivalent fraction images:(a) DECT (b)
pDECT. The second row is for bone equivalent fraction images: (c)
DECT (d) pDECT. }\label{fig6}
\end{figure}

To further evaluate the performance of the CD-ConvNet, a material
decomposition experiment is conducted.  The bone equivalent
fraction images and the water equivalent fraction images are shown
in Fig.~\ref{fig6}.  From the result, it can be seen that the
water equivalent fraction image and bone equivalent fraction image
from pDECT images are close to these from DECT images in visual
inspection.  The mean and SD values of ROI in Fig.~\ref{fig6} are
computed and the corresponding results are depicted in
Table~\ref{tab1}.

% To include a figure: 6

\begin{table}[H]\centering
\small\caption{The mean$\pm$SD measurements of ROI in
Fig.~\ref{fig6}.}
%\begin{indented}
%\item[]
\begin{tabular}{l*{3}{c c c}{C{5cm}}}
%\begin{tabular}{p{3.5cm}p{3cmp{2.5cm}}p{5cm}}
%{cccccccccc}
\hline\hline &  DECT& pDECT& \\
\hline
\multirow{1}{*}{Water equivalent fraction image}&    0.46$\pm$6.72e-8&    0.45$\pm$1.02e-8& \\
%\hline
\multirow{1}{*}{Bone equivalent fraction image}&    0.053$\pm$3.86e-8&    0.061$\pm$6.74e-9& \\

\hline\hline
\end{tabular}
%\end{indented}
\label{tab1}
\end{table}

\section{CONCLUSION}
In this work, we presented a cascade deep convolution neural
network (CD-ConvNet) structure to simulate pseudo high-energy
images from the low-energy CT images in deep learning based
framework. Clinical patient data were employed to validate and
evaluate the presented CD-ConvNet approach performance.  Both
visual and quantitative results demonstrate the presented
CD-ConvNet approach can yield high quality pDECT images.  It is
also noted that we can obtain various energy-specific pseudo DECT
images in the presented network depending on training datasets.

However, the CD-ConvNet can not produce the exact CT value
compared with reference images, which is the inherent weakness of
convolutional neural networks.  In further study, various
approaches are can be conducted to improve the performance of the
CD-ConvNet.  We also investigate other deep learning based network
for DECT imaging, i.e., generative adversarial network
(GAN)\cite{Goodfellow2014}, multi-task learning
network\cite{Evfeniou2004}.

%%%%%%%%%%%%%%%%%%%%%%%%%%%%%%%%%%%%%%%%%%%%%%%%%%%%%%%%%%%%%
\acknowledgments     %>>>> equivalent to \section*{ACKNOWLEDGMENTS}
%This work was supported in part by the National Natural Science
%Foundation of China under Grants 81371544, 61571214, 81501466 and
%81501541, the National Science and Technology Major Project of the
%Ministry of Science and Technology of China under Grant
%2014BAI17B02, Guangdong Natural Science Foundation under Grants
%2015A030313271, 2014A030310243 and 2015A030310018, and the Science
%and Technology Program of Guangzhou, China under Grant
%201510010039.
This work was supported in part by the National
Natural Science Foundation of China under Grant Nos. 81371544 and
61571214, the China Postdoctoral Science Foundation funded project
under Grants Nos. 2016M602489 and 2016M602488, the Guangdong
Natural Science Foundation under Grant Nos. 2015A030313271, the
Science and Technology Program of Guangdong, China under Grant No.
2015B020233008, the Science and Technology Program of Guangzhou,
China under Grant No. 201510010039.

%%%%%%%%%%%%%%%%%%%%%%%%%%%%%%%%%%%%%%%%%%%%%%%%%%%%%%%%%%%%%
%%%%% References %%%%%


\begin{thebibliography}{1}

%\bibitem{Graser2009}
%Graser A, Johnson T R, Hecht E M, et al., ``Dual-energy CT in
%patients suspected of having renal masses: can virtual nonenhanced
%images replace true nonenhanced images?,'' {\em Radiology}~{\bf
%252}(2),433 (2009).

%\bibitem{Danad2015}
%Danad I, Min J K., ``Dual-energy computed tomography for detection
%of coronary artery disease,'' {\em Expert Review of Cardiovascular
%Therapy}~{\bf 13}(12) 1345 (2015).

%\bibitem{Dong2016}
%Zeng D, Huang J, Zhang H, et al., ``Spectral CT Image Restoration
%via an Average Image-induced Nonlocal Means Filter,'' {\em IEEE,
%International Symposium on Biomedical Imaging}~ 497-500, (2016).

%\bibitem{Dong_2016}
%Zeng D, Gao Y, Huang J, et al., ``Penalized weighted least-squares
%approach for multienergy computed tomography image reconstruction
%via structure tensor total variation regularization,'' {\em
%Computerized Medical Imaging and Graphics}~ {\bf 53} 19-29,(2016).
%
%\bibitem{zhang_2017}
%Zhang H J,Zeng D,rt al., ``Iterative reconstruction for dual
%energy CT with an average image-induced nonlocal means
%regularization,'' {\em Physics in Medicine and Biology}~{\bf 62}
%55-56, (2017).

%\bibitem{Xe2012}
%Xie J, Xu L, Chen E., ``Image denoising and inpainting with deep
%neural networks,'' {\em International Conference on Neural
%Information Processing Systems}~341-349, (2012).

%\bibitem{He2016}
%He, K, Zhang,X, Ren,S, and Sun,J., ``Deep residual learning for
%image recognition,'' {\em In Proceedings of the IEEE Conference on
%Computer Vision and Pattern Recognition}~ 770-778, (2016).

%\bibitem{DongC2014}
%Dong C, Loy C C, He K, et al., ``Image Super-Resolution Using Deep
%Convolutional Networks,'' {\em IEEE Transactions on Pattern
%Analysis and Machine Intelligence}~{\bf 38}(2) 295-307, (2014).

%\bibitem{Glorot2010}
%Glorot, X. and Bengio, Y., ``Understanding the difficulty of
%training deep feedforward neural networks,'' {\em In Aistas}, {\bf
%9}~ 249-256, (2010).

\bibitem{Lecun2015}
Lecun Y, Bengio Y, Hinton G., ``Deep learning,'' {\em Nautre}~{\bf
521}(7553) 436-444, (2015).

\bibitem{Goodfellow2014}
Goodfellow I, Pouget-Abadie J, Mirza M, et al., ``Generative
adversarial nets,'' {\em Advances in neural information processing
systems}~ 2672-2680, (2014).

\bibitem{Evfeniou2004}
Evgeniou T, Pontil M., ``Regularized multi--task learning,'' {\em
Proceedings of the tenth ACM SIGKDD international conference on
Knowledge discovery and data mining}~ 109-117,(2004).


%\bibitem{Alred03}
%Shikhaliev, P.M., Xu, T., and Molloi, S.,  [{\em Handbook of
%Technical
%  Writing}{\nolinebreak\hspace{0.1em}]}, St. Martin's, New York (2003 (seventh
%  edition)).
%
%\bibitem{Goossens97}
%Goossens, M., Mittelbach, F., and Rahtz, S.,  [{\em The LaTeX
%  Companion}{\nolinebreak\hspace{0.1em}]}, Addison-Wesley, Reading, Mass.
%  (1997).
%
%\bibitem{Metropolis53}
%Metropolis, N., Rosenbluth, A.~W., Rosenbluth, M.~N., Teller,
%A.~H., and
%  Teller, E., ``Equations of state calculations by fast computing machine,''
%  {\em J. Chem. Phys.}~{\bf 21},  1087--1091 (1953).
%
%\bibitem{Gull89a}
%Gull, S.~F., ``Developments in maximum-entropy data analysis,'' in
%[{\em
%  Maximum Entropy and Bayesian Methods}{\nolinebreak\hspace{0.1em}]},
%  Skilling, J., ed.,  53--71, Kluwer Academic, Dordrecht (1989).
%
%\bibitem{Hanson93c}
%Hanson, K.~M., ``Introduction to {B}ayesian image analysis,'' in
%[{\em Medical
%  Imaging:\ Image Processing}{\nolinebreak\hspace{0.1em}]},  Loew, M.~H., ed.,
%  {\em Proc. SPIE} {\bf 1898},  716--731 (1993).

\end{thebibliography}
\end{document}